\documentclass[preprint]{elsarticle}

\newcommand{\del}{\partial}

\usepackage{graphicx,color,mathrsfs,flafter,amsmath,amssymb,amsfonts}
\journal{Physics Letters B}

\begin{document}
\begin{frontmatter}

\title{ Non-perturbative dynamics and charge fluctuations in  effective chiral models}

\author[GSI]{V. Skokov} \address[GSI]{%
  GSI Helmholtzzentrum f\"ur Schwerionenforschung, D-64291 Darmstadt,
  Germany}
\author[GSI]{B.~Friman}
\author[Pol,Emmi]{and K.~Redlich}

\address[Pol]{%
  Institute of Theoretical Physics, University of Wroclaw, PL--50204
  Wroc\l aw, Poland}
	\address[Emmi]{%
ExtreMe Matter Institute EMMI, GSI, D-64291 Darmstadt, Germany}

\begin{abstract}
We discuss the properties of fluctuations of the electric charge in the vicinity of the chiral crossover transition within effective chiral models at finite temperature and vanishing net baryon density. The calculation includes non-perturbative dynamics implemented within the functional renormalization group approach. We study the temperature dependence of the electric charge susceptibilities in the linear sigma model and explore the role of quantum statistics. Within the Polyakov loop extended quark-meson model, we study the influence of the coupling of quarks to mesons and to an effective gluon field on charge fluctuations. We find a clear signal for the chiral crossover transition in the fluctuations of the electric charge. Accordingly, we stress the role of higher order cumulants as probes of criticality related to the restoration of chiral symmetry and deconfinement.

\end{abstract}


\end{frontmatter}

 %\pacs{24.85.+p,21.65.-f,25.75.-q,24.60.-k}

% \date{\today}

\section{Introduction}

Thermodynamics of strongly interacting matter has been explored
numerically within  Lattice Quantum Chromodynamics (LQCD)~\cite{L_lgt1,L_lgt2,L_lgt3,Aoki:2006we}
 as well as experimentally
in  heavy-ion collisions \cite{qm}.
 The fundamental goal of these studies  is to
reveal the structure of the QCD phase diagram
and to investigate the critical properties of QCD.

LQCD results show that at finite temperatures QCD exhibits restoration of chiral symmetry
and deconfinement in a crossover transition.  The LQCD
equation of state indicates a clear separation between the confined
hadronic phase and the deconfined quark--gluon plasma.
  The temperature separation of the phases of QCD is  very transparent when one considers fluctuations of conserved charges \cite{L_lgt2,Ejiri:2005wq,Ejiri,lgt2,karsch,ratti}. At vanishing chemical potential the variance of both the net baryon number and the electric charge exhibits a rapid change in a narrow temperature interval. This behavior is attributed to the deconfinement of quarks \cite{Skokov:2011,stokic,kr}.  On the other hand, the fourth order cumulants of these charges show  peak-like structures,  whereas the sixth order cumulants are  negative in a narrow  temperature interval \cite{Ejiri:2005wq}.

These properties of the fluctuations observed in LQCD at finite temperature and for  small  masses of the  up-- and down--quarks can be attributed to the critical dynamics of the explicitly broken chiral symmetry, which is characterized by  the O(4) universality class of  QCD \cite{Ejiri:2005wq,bengt}. As noted in~\cite{bengt,prob}, this could have  interesting phenomenological implications: it has been suggested, that the sixth  and eight  order cumulants  of the net baryon number and electric charge may allow an experimental verification of the chiral crossover transition in heavy-ion collisions \cite{prob}.

The critical behavior  of strongly interacting matter related with chiral dynamics should be  common to all models which exhibit the underlying chiral symmetry of QCD and the same patterns of spontaneous chiral symmetry breaking at finite temperature.
Thus, such  effective models have been used to  study the thermodynamics near the chiral phase transition and to explore observables which are sensitive  to the critical behavior \cite{Buballa:review,Fukushima,PNJL,CS,WW,Fukushima:strong,Fu:2009wy,sasaki}.
In this context,  models  which  include the coupling of  quarks to an effective gluon field,  such as the Polyakov loop extended Nambu--Jona--Lasinio
(PNJL) \cite{PNJL,Fukushima:strong} and the quark--meson (PQM) \cite{Schaefer:PQM} models, are very  useful. Both  models  reproduce the
essential features of the QCD thermodynamics already in the mean-field approximation \cite{Buballa:review,Fukushima,PNJL,CS,Fukushima:strong}.
However, to correctly account for the critical dynamics, it is necessary  to go beyond the
mean-field approximation and include non-perturbative fluctuations.
This can be achieved e.g. by employing methods based on the functional renormalization group
(FRG)~\cite{Wetterich,Morris,Ellwanger,Berges:review}.
The FRG has been applied to study  the fluctuations of the net baryon number near the chiral phase transition at finite and at vanishing baryon chemical potential \cite{bengt,Skokov:2010wb,Skokov:2011,kr,vs}. However, so far such  studies are not available for the electric charge fluctuations.

In this work, we study fluctuations of the electric charge near the chiral phase transition at finite temperature and at vanishing chemical potential within the FRG approach to the linear sigma and PQM models.
We explore  the temperature dependence of these fluctuations and the role of quantum statistics. Furthermore, the influence of quark- and gluon-dynamics  on the properties of the electric charge fluctuations  near the chiral crossover transition, including the effect of mesonic fluctuations, is discussed.
We find a strong sensitivity of higher order cumulants  of the electric charge to chiral symmetry breaking and deconfinemenet. We also show,
that these cumulants exhibit a characteristic O(4) structure, similar to that
found in the fluctuations of the net baryon charge. In particular,
the cumulants turn negative in the vicinity of the chiral transition also at vanishing net charge.

In the subsequent section we introduce the  non-perturbative thermodynamic potential in the FRG approach to the PQM model formulated at finite electric-charge chemical potential. In section 3 we present our results on  fluctuations of the electric charge, including higher cumulants, in the linear sigma and PQM models. Finally, in section 4 we present our conclusions.

\section{The thermodynamic potential in the Polyakov-quark-meson model}\label{sec:pqm}

The quark--meson model is an effective realization of the low--energy
sector of QCD, which incorporates chiral symmetry and exhibits  a global
 $SU(N_c)$ color
symmetry. Hence,  it   does not describe quark confinement. Nevertheless, by introducing a coupling of the quarks
to a uniform temporal color gauge field,  represented by the Polyakov loop, one obtains an equation of state with properties that closely resemble those of QCD matter, including a change of effective degrees of freedom with increasing temperature, thus mimicking the confinement-deconfinement transition ~\cite{Fukushima,CS,
Fukushima:strong, Schaefer:PQM}. This is sometimes referred to a {\em statistical confinement}.

A formulation of the thermodynamics of the PQM model, which remains valid near the chiral phase transition, requires the  use of non-perturbative methods. In an analysis of electric charge fluctuations, it is of particular importance to account for mesonic fluctuations, including specifically interacting pions as carriers of electric charge.
We employ a  method  based on the functional renormalization group (FRG) to compute the thermodynamic potential in the PQM model.
This method involves an infrared regularization of the fluctuations at a sliding momentum scale
$k$, resulting in a scale-dependent
effective action $\Gamma_k$ ~\cite{Wetterich, Morris, Ellwanger, Berges:review}.
We treat the Polyakov loop
as a background field, which is introduced self-consistently on the
mean-field level while the quark and meson fields,  fluctuations are accounted for by solving the FRG flow equations.

We follow the procedure used in Ref.~\cite{Skokov:2010wb} in the formulation of the
flow equation for the scale-dependent grand canonical potential density, $\Omega_{k}=T\Gamma_{k}/V$, for
the quark and meson subsystems at finite temperature and for a non-vanishing electric charge chemical potential. The thermodynamic potential is obtained by solving the  flow equation
  \begin{eqnarray}\label{eq:frg_flow}
    &&\del_k \Omega_k(\ell, \ell^*; T,\mu)=\frac{k^4}{12\pi^2}
    \left\{ \frac{1}{E_\pi} \Bigg[ 1+2n_B(E_\pi;T)\Bigg]  \right. \nonumber \\&&+ \left. \frac{1}{E_\pi} \Bigg[ 1+2n_B(E_\pi-\mu_\pi;T)\Bigg]
 + \frac{1}{E_\pi} \Bigg[ 1+2n_B(E_\pi+ \mu_\pi;T)\Bigg] \right. \\
      &&+\frac{1}{E_\sigma} \Bigg[ 1+2n_B(E_\sigma;T)
      \Bigg]     \left. -\sum_{f=u,d} \frac{4 N_c}{E_q}  \Bigg[ 1-
      N(\ell,\ell^*;T,\mu_f)-\bar{N}(\ell,\ell^*;T,\mu_f)\Bigg] \right\}. \nonumber
  \end{eqnarray}
	  Here $n_B(E;T)$ is the bosonic distribution function
  \begin{equation*}
    n_B(E;T)=\frac{1}{\exp({E/T})-1},
  \end{equation*}
$\mu_\pi=e_\pi \mu_Q$ is the charge pion chemical potential and $e_\pi=1$  the charge of a $\pi^{+}$.
 The pion and sigma energies are given by
  \begin{equation*}
    E_\pi = \sqrt{k^2+\overline{\Omega}^{\,\prime}_k}\;~,~ E_\sigma
    =\sqrt{k^2+\overline{\Omega}^{\,\prime}_k+2\rho\,\overline{\Omega}^{\,
        \prime\prime} _k},
  \end{equation*}
where the primes denote derivatives with respect to $\rho = (\sigma^2+\vec{\pi}^2)/2$ of
  $\overline{\Omega}=\Omega+c\sigma$.
  The fermion distribution functions $N(\ell,\ell^*;T,\mu_f)$ and
  $\bar{N}(\ell,\ell^*;T,\mu_f)$,
  \begin{eqnarray}\label{n1}
    N(\ell,\ell^*;T,\mu_f)&=&\frac{1+2\ell^*e^{\beta(E_q-\mu_f)}+\ell e^{2\beta(E_q-\mu_f)}}{1+3\ell e^{2\beta(E_q-\mu_f)}+
      3\ell^*e^{\beta(E_q-\mu_f)}+e^{3\beta(E_q-\mu_f)}},  \\
    \bar{N}(\ell,\ell^*;T,\mu_f)&=&N(\ell^*,\ell;T,-\mu_f),
    \label{n2}
  \end{eqnarray}
  are modified because of the  coupling to the gluon field. Finally, the quark energy reads
  \begin{equation}
    \label{dispertion}
    E_q =\sqrt{k^2+2g^2\rho}
  \end{equation}
and the quark chemical potentials are defined by
	\begin{equation}
	\mu_u =  \frac13 \mu_B + e_u \mu_Q, \quad \mu_d =  \frac13 \mu_B + e_d \mu_Q
  \label{QChP}
	\end{equation}
	with $e_u=2/3$ and $e_d=-1/3$.

The flow equation~(\ref{eq:frg_flow}) is solved numerically with the
ultraviolet cutoff $\Lambda=1.2$ GeV using the polynomial method described in
Ref.~\cite{Skokov:2010wb}.  In this scheme, the
stationarity condition
\begin{equation}
  \left. \frac{d \Omega_k}{ d \sigma} \right|_{\sigma=\sigma_k}=\left. \frac{d
      \overline{\Omega}_k}{ d \sigma} \right|_{\sigma=\sigma_k} - c =0
  \label{eom_sigma}
\end{equation}
is implemented in the flow equation. The initial conditions for the flow are
chosen to reproduce the following  vacuum properties: the physical pion mass $m_{\pi}=138$
MeV, the pion decay constant $f_{\pi}=93$ MeV, the sigma mass
$m_{\sigma}=600$ MeV, and the constituent quark mass $m_q=300$ MeV at
the scale $k\to 0$.  The symmetry breaking term, $c=m_\pi^2 f_\pi$,
is treated as  an external field which does not flow.
The flow of the Yukawa coupling $g$ is neglected
because it is not expected to be significant for
the present studies~(see e.g. Refs.~\cite{Jungnickel}).

By solving Eq.~(\ref{eq:frg_flow}) one obtains the thermodynamic
potential
$\Omega_{k\to0} (\ell,
\ell^*;T, \mu)$ as a function of the Polyakov loop variables $\ell$
and $\ell^*$. The full thermodynamic potential $\Omega(\ell, \ell^*;T,
\mu)$ in the PQM model, including quark, meson and gluon
degrees of freedom, is obtained by adding the effective gluon potential ${\cal U}(\ell,
\ell^*)$,
\begin{equation}
  \Omega(\ell, \ell^*;T, \mu) = \Omega_{k\to0} (\ell, \ell^*;T, \mu) + {\cal U}(\ell, \ell^*).
  \label{omega_final}
\end{equation}
At a given temperature and chemical potential, the Polyakov loop
variables, $\ell$ and $\ell^*$, are then determined by the stationarity
conditions:
\begin{equation}
  \label{eom_for_PL_l}
 \frac{ \partial   }{\partial \ell} \Omega(\ell, \ell^*;T, \mu)  =0, \quad \frac{ \partial   }{\partial \ell^*}  \Omega(\ell, \ell^*;T, \mu)   =0.
\end{equation}

The thermodynamic potential~(\ref{omega_final}) does not contain
contributions of thermal modes with momenta larger than the cutoff
$\Lambda$.  In order to obtain the correct high-temperature behavior
of the thermodynamic functions, we supplement the FRG potential with the
contribution of the high-momentum states.  A  procedure for
implementing this was proposed in Ref.~\cite{Braun:2003ii} for the QM model
and extended to the PQM model in Ref.~\cite{Skokov:2010wb}, by including 
the flow of quarks interacting with the Polyakov loop for momenta $k >
\Lambda$.

\section{Electric charge density fluctuations}

The thermodynamic potential obtained by solving the flow equation (\ref{eq:frg_flow}) correctly reproduces the critical O(4) scaling near the chiral phase transition. Thus, using this potential one can compute the temperature dependence of the electric charge fluctuations and their higher cumulants.

The  fluctuations of the electric charge  are
characterized by the generalized susceptibilities,
\begin{equation}
  \chi_n^Q(T)=\frac{\del^n[p\,(T,\mu_Q)/T^4]}{\del(\mu_Q/T)^n}.\label{susceptibility}
\end{equation}
The first cumulant,  $\chi^Q_1=n_Q/T^3$, is given by 
the electric charge density $n_Q=N_{Q}/V$, while the second cumulant
\begin{equation}
  \chi_2^Q =  \frac{1}{V T^3}  \langle(\delta N_Q )^2\rangle
\end{equation}
with $\delta N_Q= N_Q-\langle N_Q\rangle$, is proportional to the variance  of the electric charge.
Furthermore, the  fourth and sixth order cumulants  can be  expressed through moments of  $\delta N_Q$ 
\begin{eqnarray}
  \chi_4^Q &=& \frac{1}{V T^3} \left[\langle(\delta N_Q)^4\rangle-3\langle(\delta N_Q)^2\rangle^2\right], \\ \label{fluctuations}
  \chi_6^Q &=& \frac{1}{V T^3} \left[    \langle(\delta N_Q)^6\rangle  -15 \langle(\delta N_Q)^4\rangle \langle(\delta N_Q)^2\rangle -10 \langle(\delta N_Q)^3\rangle^{2}\right.\\ &+& \left.30\langle(\delta N_Q)^2\rangle^3        \right].\nonumber
\end{eqnarray}

\begin{figure*}[t]
  \includegraphics*[width=6cm]{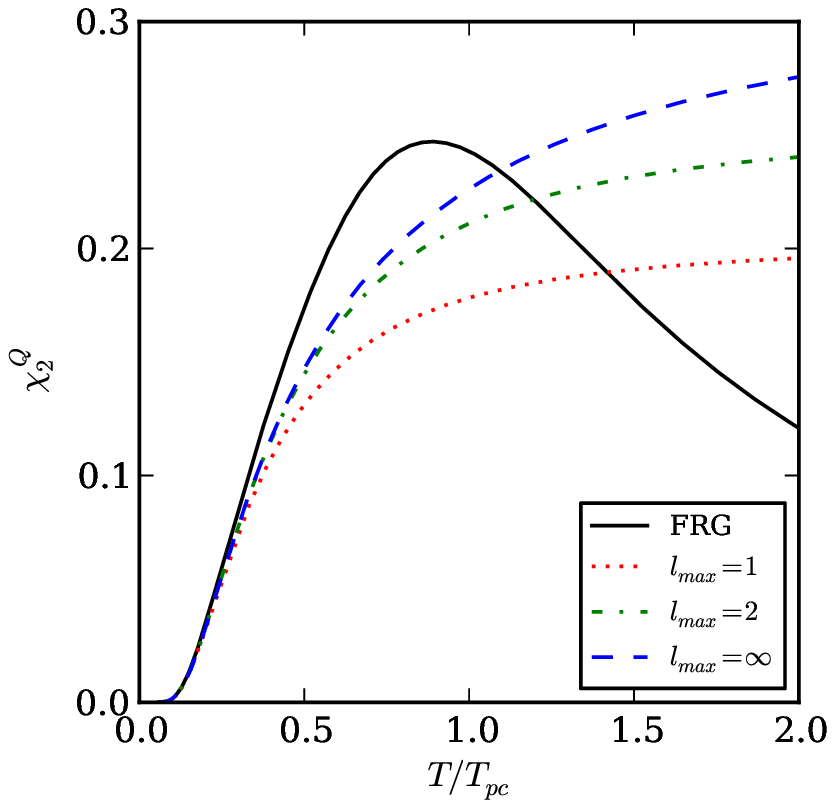}
  \includegraphics*[width=6cm]{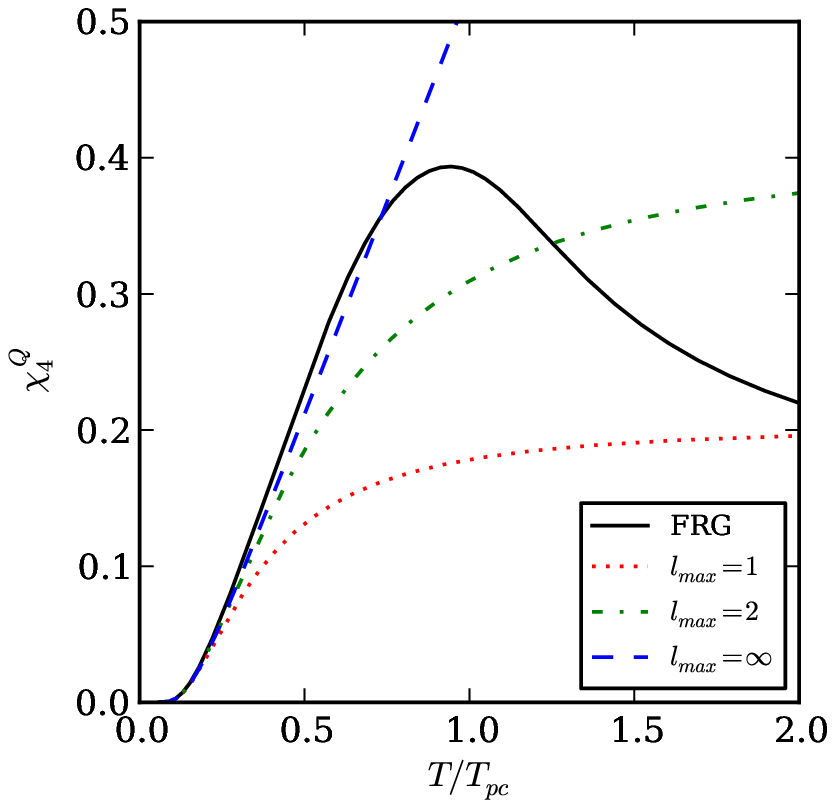}
  \caption {(Color online)  The second $\chi_2^Q$ and the fourth $\chi_4^Q$ order cumulant of the net electric charge fluctuations
	in the  linear sigma model obtained within the functional renormalisation group approach (solid lines).  
	The dotted line is the contribution of an ideal pion gas in the Boltzmann approximation. 
	The dash-dotted ($l_{max}=1$) and  dashed ($l_{max}=2$) lines show the first and the second terms in the series (\ref{chipion}), respectively.    The temperature  axis is normalized by
	$T_{pc}=180.5$ MeV which is the pseudo-critical temperature in the model within  the  FRG approach.
  \label{fig:cn_po}
	}
\end{figure*}

Based on LQCD  results \cite{karsch2}, one expects that  in the region of the
crossover transition, the fluctuations of the electric charge should reflect the critical scaling of the underlying O(4)
symmetry.
For the O(4) universality class,
the singular part of thermodynamic  pressure, at small values of the baryon chemical potential, scales   as
\begin{equation}\label{eq:sing}
p\propto (T/T_c-1 + \kappa \mu_B^2 )^{2-\alpha},
\end{equation}
where $T_c$ is the
temperature of the second-order phase transition at $\mu=0$ in the chiral limit. The exponent  $\alpha\approx-0.21$ is the critical exponent of the specific heat in the O(4) universality class in three-dimensions.
Therefore, at vanishing chemical potential, the cumulants
$\chi_n^B$ of the net baryon number  diverge at the transition temperature  for all  even $n\geq 6$, 
\begin{equation}
\chi^B_{2n} \propto (T/T_c-1)^{2-n-\alpha}.
\end{equation}

Fluctuations of the electric charge are related to the net baryon and isospin fluctuations by
\begin{equation}
\chi_n^Q = \frac{1}{2^n} \left[   \chi_n^B + \chi_n^I + \sum_{i=1}^{n-1} {{i}\choose{n}}   \frac{\partial^n (p \beta^4)}{\partial (\beta \mu_I)^{i}\partial (\beta \mu_B)^{n-i}  } \right].
\label{chi}
\end{equation}
The isovector fluctuations are regular at the chiral phase transition in an isospin symmetric system. Thus, the $\chi_n^I$ remain finite at the phase transition.   The last  term,  which correlates isospin with baryon number, can diverge for sufficiently large  $n$. However,  these terms  give subleading contributions
to criticality of $\chi_n^Q$.
Thus, the leading singular part of
the electric-charge  cumulants  $\chi_n^Q$ is solely determined  by  $\chi_n^B$.
Consequently, at small quark masses,  the   contributions from  the singular part of the thermodynamic pressure to
 the  net electric charge fluctuations  should be similar  to  that found for  the net baryon fluctuations.

The  expected residual O(4) scaling in $\chi_n^B$ at finite quark masses  is indeed observed in the PQM  model in the FRG approach. However, since the regular contributions to the electric charge and to the baryon number fluctuations are different, it is not {\em a priori} clear that this will be the case also for cumulants  $\chi_n^Q$ of the electric charge.

 In the following, we compute the temperature dependence of $\chi_n^Q$ in the PQM model with the thermodynamic potential obtained from Eq. (\ref{omega_final}). We also discuss the contributions of individual modes to the properties of $\chi_n^Q$ near the chiral transition.

	\begin{figure*}[t]
  \includegraphics*[width=3.9cm]{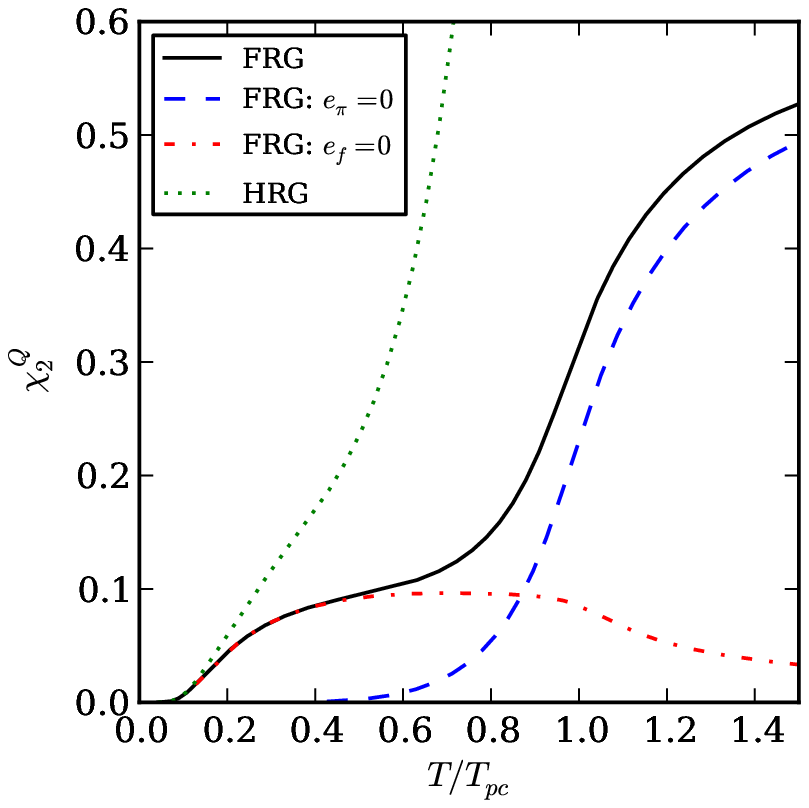}
  \includegraphics*[width=3.9cm]{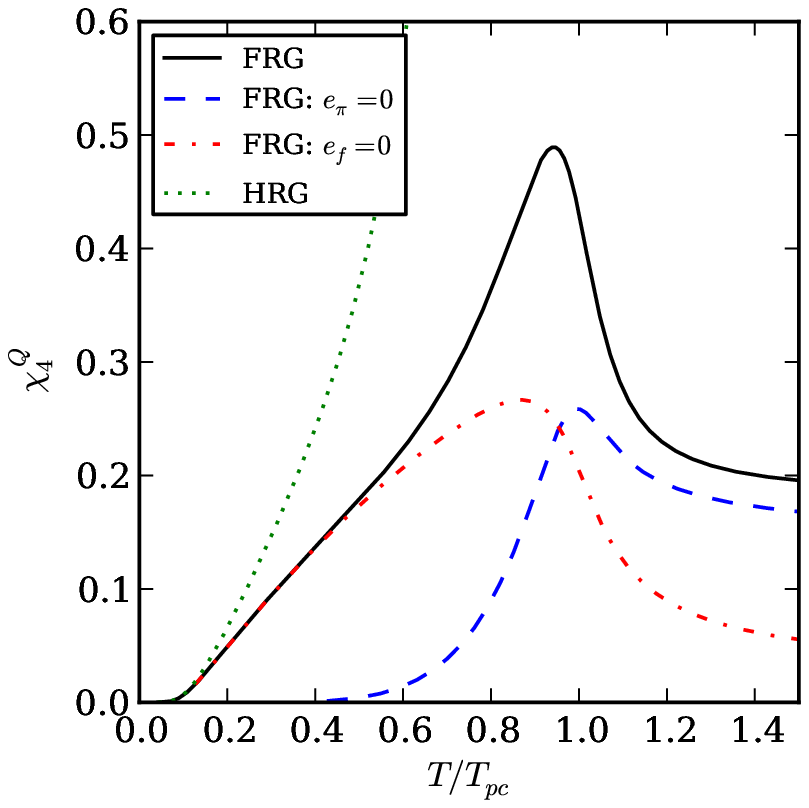}
  \includegraphics*[width=3.9cm]{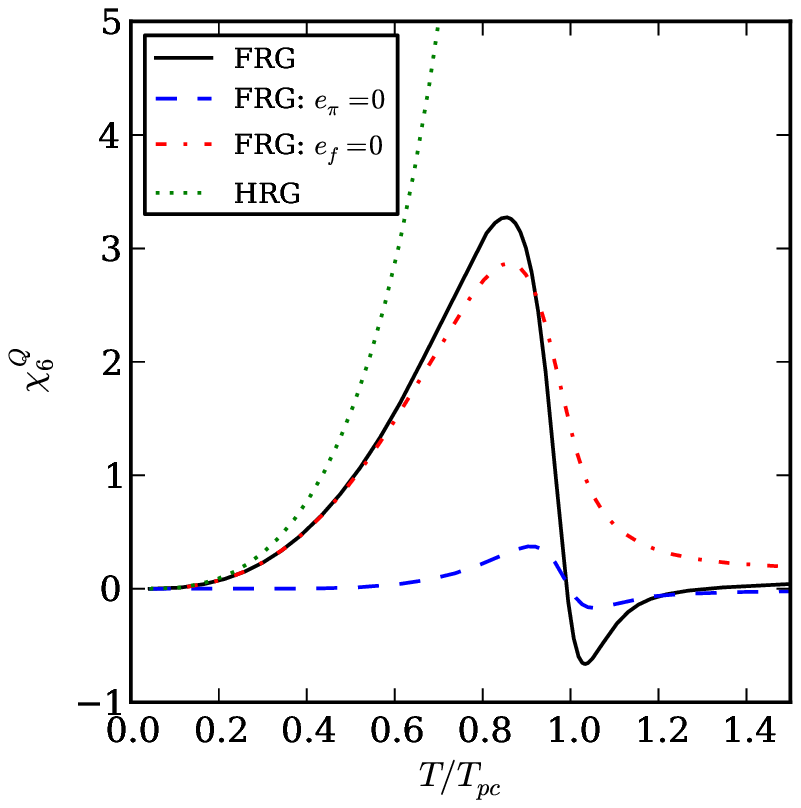}
  \caption {(Color online)  The cumulants of the electric charge fluctuations
	in the Polyakov loop extended quark-meson model calculated within the functional renormalisation group approach (solid lines). The  results are also shown for
	only charged quarks $e_\pi=0$ and only charged pions $e_f=0$ (see text). The hadron resonance gas contribution is shown as dotted lines.
  \label{fig:cn_f}
	}
\end{figure*}

\subsection{Fluctuations in an interacting pion gas}

We first consider the contribution of mesonic degrees of freedom
to  the fluctuations of the net electric charge. Clearly, in such studies pionic interactions should be included. A suitable model  for exploring the meson contribution is the  mesonic linear sigma model. The thermodynamic potential for this  model is  obtained directly from the  flow equation (\ref{omega_final})
in the limit of  $N_c\to 0$.  The fluctuations of the electric charge are then due only to the charged pion degrees of freedom.

In the mean-field approximation to  the linear sigma model, fluctuations of the electric charge are absent. They can  be included perturbatively
\cite{blaschke,blaschke1,weise} or non-perturbatively within the FRG approach, where fluctuations due to interacting pions are included \cite{Skokov:2011,stokic,Skokov:2010wb,sch1}. Here we employ the FRG approach to account for meson fluctuations.

The second and the fourth order cumulants obtained in the FRG approach are shown in  Fig. \ref{fig:cn_po}. Although the chiral symmetry is explicitly broken by the  vacuum pion mass, a clear signature of the chiral crossover transition is seen in the susceptibilities.
In particular, we note the non-monotonic temperature dependence of the fourth order cumulant, resulting in a peaked structure near the chiral transition. This behavior is due mostly to the temperature dependence of the pion mass, which up to the chiral transition remains close to its vacuum value, while above the transition $m_{\pi}$ grows strongly with temperature. Consequently, above the pseudo-critical temperature $(T_{\rm pc})$,  all cumulants of the net-charge fluctuations decrease  with temperature. Note that this effect is not a consequence of the cut off, which suppresses high momentum modes in the model. Due to the large value of  $\Lambda=1.2$ GeV, this suppression sets in only at much higher temperatures.

 It is interesting to asses at which temperature pionic interactions start influencing the charge fluctuations. This can be estimated by comparing the FRG results for the $\chi^Q_2$ and $\chi^Q_4$   with those obtained for an ideal gas of charged pions.
The pressure of the latter
is given by
\begin{equation}
\frac{p_{\pi^+} +  p_{\pi^-}}{T^4} = \frac{m^2} {2\pi^2 T^2} \sum_{l=1}^{\infty} \frac{1}{l^2} K_2 (l \beta m) \left( e^{l \beta \mu_q} +  e^{-l \beta \mu_q}  \right),
\label{ppion}
\end{equation}
where $K_2(x)$ is a modified Bessel function and $\mu_q$ is the electric chemical   potential.
In an ideal  pion gas,   the odd-order cumulants of the electric charge
vanish at $\mu_q=0$,   while even cumulants are given by
\begin{equation}
\chi^Q_n = \frac{m^2} {\pi^2 T^2} \sum_{l=1}^{\infty} l^{n-2} K_2 (l \beta m).
\label{chipion}
\end{equation}
This can be considered as an approximation to the regular part of the  fluctuations of the electric charge.

In Fig.~\ref{fig:cn_po} the contributions of the regular part to the second and fourth cumulants are shown. At  low temperatures  the FRG  and  ideal pion gas results agree. This indicates  that the FRG method correctly accounts for the pionic contribution to the thermodynamics.

The effect of quantum statistics can be studied by truncating the series in Eq. (\ref{chipion}) at a given order $l=l_{max}$.
This is illustrated in Fig.~\ref{fig:cn_po}, where we confront the $l_{max}=1$ and $2$ truncations for the second and the fourth order cumulants of the free pion gas with the full ($l_{max}=\infty$) result.
Clearly,  the leading contribution, which corresponds to Boltzmann statistics, deviates considerably from the full quantum statistics results, except at very low temperatures. 
The higher-order terms in the series contribute effectively as multiply charged particles. Hence, their contribution increases with the order of the cumulant. This expectation is confirmed by Fig.~\ref{fig:cn_po}, which shows that the deviation is stronger for $\chi_4^Q$ than for $\chi_2^Q$.
Thus, the effect of quantum statistics cannot be neglected in the calculation of fluctuations of the electric charge. This is a reflection of the fact  that multi-charged particles can yield a dominant contribution to the fluctuations of conserved charges \cite{Ejiri:2005wq}. In QCD this is the case  for fluctuations of the electric charge and strangeness, where double- and triple-charged baryons play an important role.

\begin{figure*}[t]
  {\includegraphics*[width=5cm]{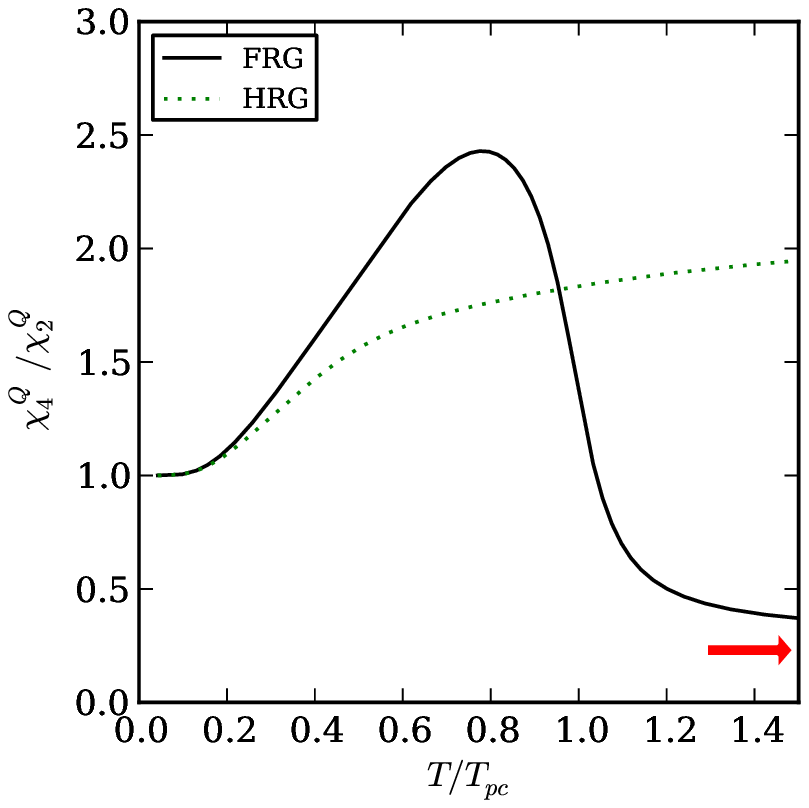}}
 \hskip 1.0cm {\includegraphics*[width=5cm]{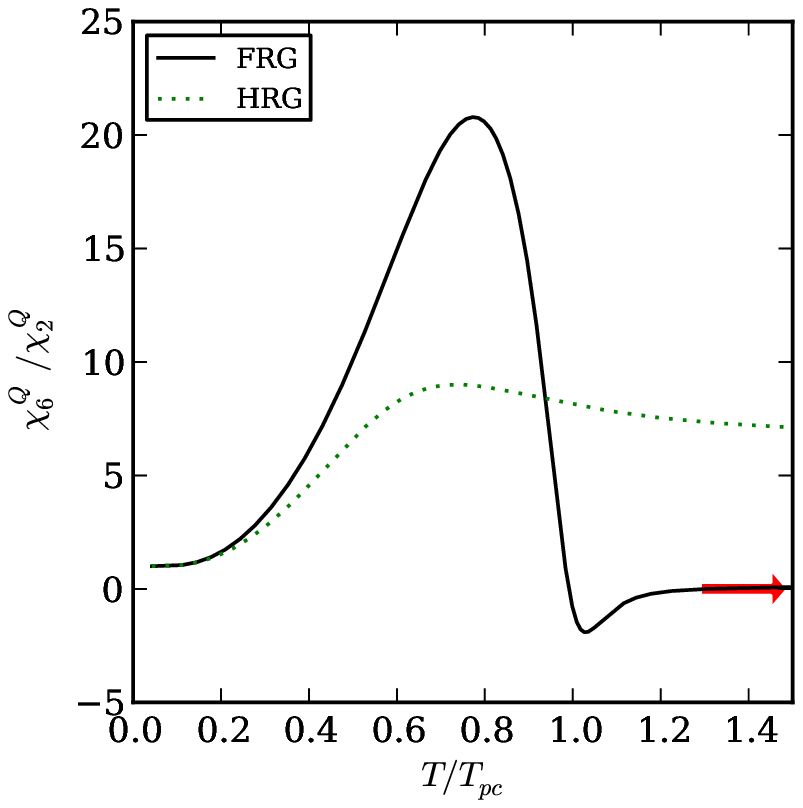}}
  \caption {(Color online)  The kurtosis ($\chi_4^Q/\chi_2^Q$) and the ($\chi_4^Q/\chi_2^Q$) ratio calculated in the Polyakov loop extended quark-meson model.  The arrow shows the corresponding Stefan-Boltzmann limits.
  The hadron resonance gas model results are  indicated as dotted-lines.
  \label{fig:k_f}
	}
\end{figure*}

\subsection{Electric charge cumulants in the PQM model}

The dynamics of the PQM model and its particle content is more relevant for QCD  than the mesonic linear sigma model described above.   The PQM model  contains  charged-quarks as dynamical degrees of freedom  and exhibits statistical confinement through the coupling of the quarks to the  effective gluon field.

In Fig.~\ref{fig:cn_f} we show the first three non-trivial cumulants $\chi_2^Q$, $\chi_4^Q$ and  $\chi_6^Q$ of the electric charge for vanishing  baryon- and electric charge chemical potential obtained  from  Eq. (\ref{omega_final}).
We also show separately the contributions of quarks or pions.
This separation is made by setting the electric charge of  pions $e_\pi$ and then that of quarks $e_f$ to zero, respectively.

The second-order cumulant $\chi_2^Q$ shows a rapid increase near the pseudo-critical temperature, interpolating between the pion and quark contributions. Since $\chi_2^Q$  is not influenced by the singular part of the thermodynamic pressure (\ref{eq:sing}), the behavior seen in Fig. \ref{fig:cn_f} is due to the ``statistical confinement'' of the PQM model. The rapid increase with temperature is a consequence of the rapid unleashing of single- and double-quarks states at the crossover temperature. The pion contribution, on the other hand, is  suppressed  above $T_{\rm pc}$, owing to the increasing thermal mass.

The fourth- and sixth-order  cumulants exhibit a peak, which increases in strength with the order of the cumulant. This can be understood in terms of the critical dynamics of the chiral transition; with increasing order of the cumulant, the contribution of the singular part of the pressure (\ref{eq:sing}) becomes more and more dominant. The negative structure of the  $\chi_6^Q$ near $T_{\rm pc}$ is similar to that observed for the sixth  order cumulant  of net baryon number fluctuations and is  due to the particular form of the O(4) scaling function \cite{bengt}. The peak in  $\chi_4^Q$ appears as a sum   of quark and pion contributions, while the peak in $\chi_6^Q$  is dominantly due  to pions.   The negative region of $\chi_6^Q$ near $T_{\rm pc}$ is due to the quark contribution.

In  Fig. \ref{fig:cn_f} we also compare the PQM  results with that of the   hadron resonance gas (HRG) model, which includes the contributions of  all charged hadrons and resonances.  The HRG model reproduces the thermodynamics of LQCD in the hadronic phase up to $T\simeq 0.9 T_{\rm pc}$.

There is good agreement between the  PQM and HRG model results at  low temperatures, where in both cases pions are the dominant degrees of freedom. For higher temperatures  the HRG  overshoots the PQM model results.  Thus, the PQM  model does not yield a quantitative description of the  LQCD results on thermodynamics. Nevertheless,  this model  can provide useful insights into the critical dynamics of universal quantities near the chiral phase transition.

In order to reduce the contribution of the non-singular part to the fluctuations and to focus on the critical behavior, ratios of $\chi_n^Q$ to the second-order cumulant $\chi_2^Q$, which is not influenced by the critical chiral dynamics at   $\mu=0$ \cite{Ejiri:2005wq}, have been studied. In Fig.~\ref{fig:k_f}  we show the temperature dependence of the  kurtosis   $\kappa=\chi_4^Q/\chi_2^Q$ and of the ratio $\chi_6^Q/\chi_2^Q$  near the chiral crossover transition. At low temperatures,  $\kappa\to 1$,  as expected for a non-interacting Boltzmann gas. For higher temperatures,  the kurtosis increases owing to the  contribution of double-charged baryons and to higher order quantum corrections to the pion distribution.
Near $T_{\rm pc}$  and above the    quarks  dominate, because of the increasing pion  mass,  resulting in    $\kappa$  decreasing towards  the Stefan-Boltzmann limit of free quark gas. The ratio $\chi_6^Q/\chi_2^Q$ exhibits a much  stronger peak, 
which diverges at the critical temperature in the chiral limit. In analogy with the net baryon fluctuations~\cite{bengt,prob}, this ratio also develops a negative region near $T_{\rm pc}$, owing to the chiral critical dynamics encoded in the O(4) scaling function.

As shown in Fig. \ref{fig:cn_f}, in the PQM model both ratios $\chi_4^Q/\chi_2^Q$ and $\chi_6^Q/\chi_2^Q$ overshoot the HRG model predictions. This indicates that in the HRG model,  the contribution of multi-charged states to fluctuations is reduced relative to that of singly-charged particles more than in the PQM model.

\section{Summary and conclusions}

We have computed the electric charge fluctuations in the vicinity of the chiral crossover transition within effective chiral models.
 Our  calculations include non-perturbative dynamics implemented within the functional renormalization group approach   at finite temperature and at vanishing baryon density. We have performed our studies in the linear sigma model and in the Polyakov--loop extended quark-meson model.

We find  a strong
   sensitivity of the fluctuations of the electric charge 
  to the chiral transition. We have also explored the influence of multi-charged states and quantum statistics on  the temperature dependence of cumulants of the electric charge.

   Our results show that at physical pion and quark masses, the higher order cumulants of the electric charge fluctuations exhibit properties which can be linked to the  chiral critical dynamics expected in the O(4) universality class. The qualitative structure of fluctuations in the PQM model is similar to that obtained in lattice QCD. This confirms the phenomenological importance of electric charge fluctuations, in particular the sixth- and higher-order cumulants,  as signatures for the chiral crossover transition  in heavy-ion collisions.

\section*{Acknowledgments}
We acknowledge stimulating discussions with Frithjof Karsch, Kenji Morita and Christian Schmidt.
V.~Skokov acknowledges the support by the Frankfurt Institute for
Advanced Studies (FIAS).  K. Redlich acknowledges partial support from the Polish
Ministry of Science (MEN). B. Friman acknowledges partial support by EMMI.

%%%%%%%%%%%%%%%%%%%%%%%%%%%%%%%%%%%%%%%%%%%%%%%%

\end{document}